\documentstyle[12pt]{article}
\def\SO{SO^{\uparrow}(3,2)}
\def\pa{\partial}
\def\cG{{\cal G}}

\def\dx{{\dot x}}

\begin{document}
\begin{titlepage}

\begin{flushright}
ITP-UH 24-95 \\
hep-th/9509062 \\
\end{flushright}

\begin{flushleft}
September 1995
\end{flushleft}

\begin{center}
\Large{{\bf Massive spinning particle on anti-de Sitter space} } \\
\vspace{1.0cm}

\large{ S.M. Kuzenko\footnote{
On leave of absence from: Department of Physics, Tomsk State
University, Tomsk 634050, Russia (address after January 1,
1996).}${}^,$\footnote{E-mail: kuzenko@itp.uni-hannover.de.}	   } \\

\footnotesize{{\it Institut f\"ur Theoretische Physik,
Universit\"at Hannover \\
Appelstr. 2, 30167 Hannover, Germany}	} \\
\vspace{0.5cm}

\large{ S.L. Lyakhovich$^{\dag,\ddag,}$\footnote{E-mail:
sll@phys.tsu.tomsk.su.},
A.Yu. Segal$^{\dag,}$\footnote{E-mail:
segal@phys.tsu.tomsk.su.} and A.A. Sharapov$^{\dag,}$\footnote{E-mail:
sharapov@phys.tsu.tomsk.su.}	}   \\

\footnotesize{{\it $^{\dag}$ Department of Physics, Tomsk State University \\
Tomsk 634050, Russia}} \\
\vspace{0.3cm}

\footnotesize{{\it $^{\ddag}$ Physics Department,
Queen Mary and Westfield College \\
Mile End Road, London E1 4NS, UK      }     }
\end{center}
\vspace{1.5cm}

\begin{abstract}
To describe a massive particle with fixed, but arbitrary, spin on $d=4$
anti-de Sitter space $M^4$, we propose the point-particle model with
configuration space ${\cal M}^6 = M^{4}\times S^{2}$, where
the sphere $S^2$ corresponds to the spin degrees of freedom.
The model possesses two gauge symmetries expressing strong conservation
of the phase-space counterparts of the second- and fourth-order Casimir
operators for $so(3,2)$. We prove that the requirement of energy to have
a global positive minimum $E_o$ over the configuration space is equivalent to
the relation $E_o > s$, $s$ being the particle's spin, what presents
the classical counterpart of the quantum massive condition.
States with the minimal energy are studied in detail. The model is
shown to be exactly solvable. It can be straightforwardly generalized
to describe a spinning particle on $d$-dimensional anti-de Sitter
space $M^d$, with ${\cal M}^{2(d-1)} = M^d \times S^{(d-2)}$
the corresponding configuration space.
\end{abstract}
\vfill
\null
\end{titlepage}

\newpage

\section{Introduction}

Not long ago, there were constructed twistor formulations for the massless
Brink-Schwarz superparticle in dimensions $d = 3,4,6$ and 10 [1,2] and later
[3] for the heterotic $d =10$ Green-Schwarz superstring, which possess
manifest invariance under both target-space supersymmetry and a world-line
(world-sheet) general covariance supergroup and provide a geometric origion
for Siegel's $\kappa$-symmetry. A central point in setting up these
formulations was the use of bosonic twistor-like variables parametrizing the
sphere $S^{d-2}$ regarded as a homogeneous space of the {\it d}-dimensional
Lorentz group [4,5].

In a recent paper [6], we proposed the model for a
massive particle of
arbitrary spin in $d = 4$ Minkowski space ${\bf R}^{3,1}$ as a
Poincar\'e-invariant
dynamical system on ${\bf R}^{3,1} \times S^2$, where $S^2$ is the
space of spin degrees of freedom. The model is based on simple physical and
geometrical principles. Its quantization leads to the unitary massive
representations of the Poincar\'e group. The physical wave-functions for
particles of all integer spins span the space of on-shell scalar fields on
${\bf R}^{3,1} \times S^2$. For a given spin, the wave-functions form an
eigenspace of a relativistic spherical Laplacian. Our model admits natural
higher-dimensional and supersymmetric generalizations [7], with
${\bf R}^{d-1,1} \times S^{d-2}$ being the bosonic part of the configuration
space.

The results of Refs. [1--3] and [6,7] indicate that extended space-times
of the form
\begin{center}
{\it d-dimensional space-time} $\times S^{d-2}$
\end{center}
deserve serious study. Their use may be relevant not only in the
superstring context but also for constructing a
consistent theory of higher-spin fields. It is worth noting that
${\bf R}^{d-1,1} \times S^{d-2}$ is the minimal homogeneous space of the
$d$-dimensional Poincar\'e group, which contains the Minkowski space as a
submanifold. In a curved space, on the other hand, the local Lorentz group
can be naturally identified with a localized version of the conformal
group of $S^{d-2}$. Manifolds of the above structure arise most
simply in the framework of a massless spinless dynamics [2].
Below we are going to show the relevance of such manifolds for describing
a massive  spin dynamics in the case of space-times with constant curvature.

In the present paper, we generalize the model of Ref. [6] to the case of
$d = 4$ anti-de Sitter (AdS) space and discuss higher-dimensional extensions.
The AdS space is known to be a maximally
symmetric solution to the Einstein vacuum equations with a negative
cosmological term (see, e.g., [8]). There are three basic grounds
in favour of the AdS space as compared to the de Sitter space (positive
cosmological constant). (i) The symmetry algebra of the AdS space, $so$(3,2),
has unitary representations with bounded energy [9--11]. The positive-energy
irreps, denoted $D(E_o,s)$, are classified by minimal energy and $E_o$ and
spin $s$, $s = 0,\frac{1}{2},1,\ldots$, with $E_o$ restricted by
unitarity as follows (in dimensionless units for energy):
$$
E_o \ge s + \frac{1}{2}, \,\,\, s =0,\frac{1}{2}\;\;\;\;\;\;\;\;\;\;\;\;\;\;
E_o \ge s + 1, \,\,\, s \ge 1.
$$
$D(\frac{1}{2},0)$ and $D(1,\frac{1}{2})$ correspond to the Dirac singletons
[12], the irreps $D(2,0)$ and $D(s + 1,s)$, for any $s$, describe massless
particles [11,13]; finally, massive particles are associated with the
choice $E_o > s + 1$.
(ii) Similarly to Minkowski space,
the AdS space can be supersymmetrized. It presents itself the even part
of AdS superspace [14,15] that originates as a maximally symmetric solution
of the superfield dynamical equatons in $N = 1, d = 4$ minimal supergravity
with a cosmological term (see, e.g., [16,17]). (iii) This is the AdS
space, neither flat nor de Sitter ones, that can arise as a classical
vacuum solution in consistent theories of higher-spin massless fields
including gravity [18,19].

To describe a spinning particle on the AdS space denoted below $M^4$,
we consider manifold ${\cal M}^{6} = M^{4} \times S^2$ as the
configuration space. ${\cal M}^{6}$ turns out to be a coset space
of the AdS group,
i.e. the symmetry group of $M^4$. The chief dynamical
principle underlying our model is the requirement of strong
conservation
for classical phase-space counterparts of the Casimir operators of the
AdS group. This principle leads to  unique gauge-invariant action functional
over ${\cal M}^{6}$. What is more, it implies automatic fulfilment of the
classical version $E_o > s$ for the quantum massive
condition  $E_o > s + 1,$ for $s \ge 1$, which was postulated in Ref. [20]
in the framework of a realization of the AdS group as a curved phase
space. Namely, by construction, the theory is
characterized by two parameters $\Delta_1$ and $\Delta_2$ which are the
values of the second- and fourth-order Casimir phase-space functions,
respectively, and can be algebraically reexpressed via some
auxiliary parameters
$E_o$ and {\it s}. Then it turns out that, first,
the energy is positive definite
over the phase space if and only if $E_o \ge s$; second, the energy possesses
a global minimum when $E_o > s$. In the latter case, $E_o$ is the
minimal value of the energy, while {\it s} coincides with the total
angular momentum at any phase-space point with the minimal energy.
Therefore, the condition $E_o > s$  specifies massive
spinning particles on the AdS space.

Another remarkable feature of the model is the fact that any physical
observable, i.e. a gauge-invariant scalar field over the phase space, proves
to be a function of the Hamilton generators of the AdS group only.
As a result, the covariant quantization of the model is equivalent to
realizing the unitary massive representations
of the AdS group in function spaces
over ${\cal M}^6$.

The paper is organized as follows. In section 2 we consider an AdS-covariant
parametrization of ${\cal M}^6$ shown to be  a homogeneous space of the
AdS group. In section 3 the action functional of the model is derived, in an
AdS-covariant way, and its local  invariances are discussed both in the first-
and second-order approaches. The main results of sections 2 and 3 can be
easily extended to the cases of $d=3$ and higher dimensions, with
${\cal M}^{2(d-1)} = M^{d} \times S^{d-2}$ being the configuration space, where
$M^d$ is a $d$-dimensional AdS space. In section 4 we investigate
the conditions under which the energy is positive definite over the
phase space and possesses a global positive minimum.
Section 5 is devoted to the description of the
model in terms of the inner ${\cal M}^6$--geometry. Here we also consider
some obstructions to a straightforward generalization of our model to the
case of arbitrary curved background. Dynamical histories with the minimal
energy are studied in section 7.
In conclusion we discuss the results
and some perspectives.

\section{Covariant realization for the configuration space}

We start with describing a covariant realization for the
configuration space ${\cal M}^{6}={M^4}\times S^2$, where
${M^4}$ presents itself an ordinary anti-de Sitter space,
$S^2$ a two-dimensional sphere. It is useful for us to treat
${M^4}$ as a hyperboloid embedded into a five-dimensional
pseudo-Euclidean space ${\bf R}^{3,2}$, with coordinates $y^A$, $A=5,
0, 1, 2, 3$, and defined by
$$
\eta_{AB}y^Ay^B=-r^2 \qquad \eta_{AB}=diag(--+++)
\eqno{(1)}$$
${\cal R}=-12r^{-2}$ is the curvature of the AdS space.
$M^4$ has the topology $S^1 \times {\bf R}^3$ and, as
a consequence, possesses closed timelike geodesics. That is why
the universal covering space $\widetilde{M}^4$ of $M^4$ is standardly chosen
to be the genuine AdS space. However, for studying the local physical
properties, that we will be mainly interested in, one can equally well
make use of $M^4$ as a model space. All our
subsequent results are readily extended to the case of $\widetilde{M}^4$.

Similarly to ${M^4}$, ${\cal M}^6$ can be endowed with the structure of a
homogeneous transformation space for an AdS group chosen below to be
the connected component of the identity in $O(3,2)$ and denoted by
$\SO$ (the elements of $\SO$ are specified
by the conditions that their diagonal $2\times 2$ and $3\times 3$
submatrices, numbering by indices 5,0 and 1,2,3,
respectively, have positive determinants). In order to explain this
statement, let us consider the tangent bundle $T({M^4})$ that will be
parametrized by 5-vector variables $(y^A,b^A)$ under the constraints
$$
y^Ay_A=-r^2
\eqno{(2.a)}$$
$$
y^Ab_A=0.
\eqno{(2.b)}$$
The latter requirement simply expresses the fact that $b^A\pa/\pa y^A$
is a tangent vector to the point $y\in{M^4}$. Embedded into $T({M^4})$
is the $O(3,2)$-invariant
subbundle $T_l({M^4})$ of non-zero lightlike tangent vectors
$$
b^Ab_A=0
\eqno{(3.a)}$$
$$
\{b^A\}\neq 0.
\eqno{(3.b)}$$
It turns out that ${\cal M}^6$ can be identified with the factor-space of
$T_l({M^4})$ with respect to the equivalence relation
$$
b^A\sim\lambda b^A \qquad \forall\lambda\in{\bf R}\setminus\{0\}.
\eqno{(4)}$$
Really, there always exists a smooth mapping
$$
{\cal G}: {M^4}\to \SO
\eqno{(5)}$$
such that ${\cal G}(y)$ moves a point $(y,b)$ at $T_l({M^4})$ to
$({\bf y,b})$ having the form
$$
{\bf y}{}^A = {{\cal G}^A}_B(y)y^B=(r,0,0,0,0)
\eqno{(6)}$$
and
$$
{\bf b}{}^A = {{\cal G}^A}_B(y)b^B=(0,{\bf b}^a) \qquad
a=0,1,2,3
\eqno{(7.a)}$$
$$
\eta_{ab}{\bf b}^a{\bf b}^b=0
\eqno{(7.b)}$$
$$
\{{\bf b}^a\}\neq 0.
\eqno{(7.c)}$$
For example, one can choose
$$
{\cal G}(y)=\left(\begin{array}{ccc}\begin{array}{cc}
\frac{\displaystyle{y^5}}{\displaystyle{r}} & \frac{\displaystyle{y^0}}
{\displaystyle{r}}\\
-~\frac{\displaystyle{y^0}}{\displaystyle\rho} & \frac{\displaystyle
y^5}{\displaystyle\rho}\end{array} &\stackrel{\vdots}{\vdots}&
\begin{array}{rrr}
-~\frac{\displaystyle y^1}{\displaystyle r} & -~\frac{\displaystyle
y^2}{\displaystyle r} & -~\frac{\displaystyle y^3}{\displaystyle r}\\
{\displaystyle 0} & {\displaystyle 0} & {\displaystyle 0}\end{array}\\
\dotfill &\vdots& \dotfill\\
\begin{array}[c]{cc}-~\frac{\displaystyle y^1y^5}{\displaystyle r\rho}
& -~\frac{\displaystyle y^1y^0}{\displaystyle r\rho}\\
-~\frac{\displaystyle y^2y^5}{\displaystyle r\rho} &
-~\frac{\displaystyle y^2y^0}{\displaystyle r\rho}\\
-~\frac{\displaystyle y^3y^5}{\displaystyle r\rho} &
-~\frac{\displaystyle y^3y^0}{\displaystyle r\rho}\end{array} &\vdots&
{\displaystyle\delta^{ij}}+\frac{\displaystyle y^iy^j}{\displaystyle
r(\rho +r)}\end{array}\right)
\eqno{(8)}$$
where
$$
\rho =\sqrt{(y^5)^2+(y^0)^2} \qquad i,j=1,2,3.
$$
${}$From Eq. (7) we see that the fiber $\{({\bf y,b})\}$ over $\bf y$ in
$T_l({M^4})$ looks exactly like the punctured light-cone in
Minkowski space. The equivalence relation (4) proves to reduce the
light-cone to $S^2$.
Now, since the AdS group brings any equivalent
points to equivalent ones, we conclude that $SO^{\uparrow}(3,2)$
naturally acts on the factor-space ${M^4}\times S^2$. Therefore, Eqs.
(2--4) present an AdS-covariant realization of ${\cal M}^6$.

There exists some inherent arbitrariness in the choice of $\cal G$
defined by Eqs. (5) and (6). Such a mapping can be equally well
replaced by another one
$$
{{\cal G}'}^A{}_B(y)={\Lambda^A}_C(y){{\cal G}^C}_B(y)
\eqno{(9.a)}$$
where $\Lambda$ takes it values in the stability group of the marked
point $\bf y$
$$
{\Lambda^A}_B(y){\bf y}{}^B={\bf y}{}^A
\eqno{(9.b)}$$
and has the general structure
$$
\begin{array}{l}
\Lambda : {M^4}\to\SO\\
{\Lambda^A}_B(y)=\left(\begin{array}{ccc} \displaystyle 1 & \vdots & 0\\
\dotfill & \vdots & \dotfill\\
\displaystyle 0 & \vdots & {\Lambda^a}_b(y)\end{array}\right) \qquad
{\Lambda^a}_b(y)\in SO^{\uparrow}(3,1).\end{array}
\eqno{(10)}$$
The set of all smooth mappings (10) forms an infinite-dimensional group
isomorphic to a local Lorentz group of the AdS space. This group acts
on $T({M^4})$ by the law
$$
(y,b)\longrightarrow\big(y,{\cal G}^{-1}(y)\Lambda(y){\cal G}(y)b\big)
\eqno{(11)}$$
$\cal G$ being a fixed solution of Eqs. (5), (6). As it
is obvious, the
local Lorentz group naturally acts on ${\cal M}^6$.

For writing down the explicit action of
$\SO$ on ${\cal M}^6$, it appears useful from
the very beginning to replace the AdS-covariant parametrization
$(y^A,b^A)$ of $T({M^4})$ with a Lorentz-covariant one $(y^A,{\bf b}^a)$, where
the 4-vector ${\bf b}^a$ is related to $b^A$ as in
Eq. (7.a). Given a group element $g\in\SO$, it moves $(y,b)$ to
$(gy,gb)$, hence $(y,{\bf b})$ to $(gy,\Lambda_g(y){\bf b})$, where
$$
\Lambda_g(y)\equiv{\cal G}(gy)g{\cal G}^{-1}(y)
\eqno{(12)}$$
is a Lorentz transformation of the form (10). One readily finds
$$
\Lambda_{g_1}(y)\Lambda_{g_2}(y)=\Lambda_{g_1g_2}(y)
\eqno{(13)}$$
for arbitrary $g_1,g_2\in\SO$. We thus arrive at a nonlinear
representation of the AdS group. Now, the problem simly reduces
to making use of the known
action of the Lorentz group on the light-cone (7.a--c).

The covariant realization of ${\cal M}^6$ described is based on the use
of a lightlike vector variable to parametrize $S^2$. Another realization,
which involves a constrained spinor variable and appears to be most suited
for constructing generalized AdS-coherent states, will be given in a
forthcoming publication [21].

\section{Derivation of the action functional}

We set about deriving the action functional
of a point particle on ${\cal M}^6$.
Our main dynamical principle is the requirement of strong
conservation for
classical counterparts of the Casimir operators of $ so(3,2) $. Let us
recall that these operators can be chosen (see, for example, [11]) in
the form
$$
\hat{\Omega}_1 =\frac{1}{2}\hat{L}^{AB} \hat{L}_{AB}
\eqno{(14.a)}$$
$$\hat{\Omega}_2 =\hat{W}^A \hat{W}_A
\eqno{(14.b)}$$
where $ \hat{W}^A $ is the Pauli-Lubanski vector
$$
\hat{W}^A =\frac{1}{8}\varepsilon^{ABCDE} \hat{L}_{BC} \hat{L}_{DE}
\eqno{(15)}$$
$\hat{L}_{AB}$ being the generators of $ so(3,2) $.

We start with an extended phase space parametrized by unconstrained
coordinates $y^A$, $b^A$ and their canonical momenta $p_A$, $k_A$
with respect to the standard
Poisson structure
$$
\{ y^A, p_B \} = \{ b^A, k_B \} =\delta^A_B
\eqno{(16)}$$
other brackets vanish. Obviously, the AdS group acts on the phase
space by canonical transformations. This action induces special
representation of this group in the space of smooth functions on
the phase space. For an infinitesimal group transformation
$$
\delta y^A = \theta ^A \,_B y^B\;\;\;\;\;\;\;\;\theta _A \,_B =-\theta _B\,_A
$$
the corresponding change of a phase-space function $\Phi(y,b,p,k)$
is given in terms of the Poisson bracket as follows
$$
\delta \Phi = \frac{1}{2}\theta^{AB} \{ {\cal J}_{AB} , \Phi \}
$$
where the Hamilton generators read
$${\cal J}_{AB} = p_A y_B + k_A b_B - (A\leftrightarrow B).
\eqno{(17)}$$

Next we introduce two sets of AdS invariant phase-space constraints:
kinematical
$$
T_1 =y^2 + r^2
\eqno{(18.a)}$$
$$
T_2 =(y,p)
\eqno{(18.b)}$$
$$
T_3 =(y,b)
\eqno{(19.a)}$$
$$
T_4 =(y,k)
\eqno{(19.b)}$$
$$
T_5 =b^2
\eqno{(20)}$$
$$
T_6 =(b,k)
\eqno{(21)}$$
and dynamical
$$
-T_7 =\Omega_1 -\Delta_1
\eqno{(22)}$$
$$
T_8 =\Omega_2 -\Delta_2.
\eqno{(23)}$$
Here $\Delta_1$ and $ \Delta_2$ are constant parameters. Evolution of the
mechanical system is postulated to develop on the full surface of constraints
$$
T_I \approx 0\;\;\;\;\;\;\;\;\;\;\;\;\;\; I=1,\ldots ,8.
\eqno{(24)}$$

Let us shortly discuss the structure and physical origion of the constraints.
The kinematical constraints are introduced in order to define $6+6$
dimensional phase space over ${\cal M}^6$, while the dynamical ones lead to
unique choice of the action functional. The constraints $T_1,\ldots ,T_4$
determine the cotangent bundle of $T({M^4})$, and $T_6$ generates the
equivalence relation (4). With respect to the Poisson bracket (16), the full
set of constraints is decomposed into two groups: second-class constraints
$T_1,\ldots ,T_4$
$$
\frac{1}{2} \{ T_1,T_2 \} = \{ T_3,T_4 \} =-r^2
\eqno{(25)}$$
and first-class constraints $T_5,\ldots ,T_8$.
The constraints $T_1,\ldots ,T_4$
can be eliminated by means of introducing local independent coordinates on
${M^4}$. To keep the AdS- covariance explicitly, however, we prefer to use
the constrained variables. It is worth noting that the constraints $T_2$ and
$T_4$ can be treated as a particular gauge fixing for the set of first-class
constraints $T_1, T_3, T_5,\ldots ,T_8$. Our subsequent results could be
obtained without imposing $T_2$ and $T_4$, but the use of these constraints
leads to maximally simple expressions for the classical counterparts of
the operators (14)
$$
\Omega_1 \approx -r^2 p^2\;\;\;\;\;\;\;\;\;\;\Omega_2 \approx r^2 (b,p)^2 k^2.
\eqno{(26)}$$

Assuming the theory to be reparametrization invariant, the Hamiltonian
is a linear combination of the constraints and the first-order (Hamilton)
action	reads
$$
S=\int{\rm d}\tau \left(p_A\dot y^A+k_A\dot b{}^A-\frac{1}{2}
\sum_{I=1}^8 \lambda_I T_I\right).
\eqno{(27)}$$
Here $\tau$ is the evolution parameter, $\lambda$'s are Lagrange multipliers
associated to the constraints. The action can be readily brought to a
second-order (Lagrange) form by eliminating the momenta $p_A $, $k_A$ and
kinematical multipliers $\lambda_1 ,\ldots , \lambda_6$ with the aid of their
equation of motion
$$
\frac{\delta S}{\delta p_A}= \frac{\delta S}{\delta k_A}=
\frac{\delta S}{\delta\lambda_i}=0\;\;\;\;\;\;\;\;\;\;\;\;\;\; i=1, \ldots , 6.
\eqno{(28)}$$
This leads to
$$
S=\int{\rm d}\tau{\cal L}
$$
$$
{\cal L} = \frac{1}{2e_1}\big({\dot y}^2-\frac{\Delta_1}{r^2}
e_1^2\big)+\frac{1}{2e_2}\bigg(\Big(\frac{{\dot b}^2}{(\dot y , b)^2}
+\frac{1}{r^2}\Big)e_1^2+\frac{\Delta_2}{r^2}e_2^2\bigg).
\eqno{(29)}$$
Here $e_1 \equiv r^2\lambda_7 $, $e_2 \equiv r^2\lambda_8 $ and the variables
$y^A$ and $b^A$ are restricted to satisfy the holonomic constraints (18.a),
(19.a) and (20).

The first-order action possesses four gauge invariances associated
with the first-class constraints $T_5,\ldots ,T_8$. After passing to the
second-order action, there remain only three gauge symmetries related to
$T_6,T_7$ and $T_8$. Each gauge transformation can be represented as a
combination  of some reparametrization of the world line
$$
\delta_{\epsilon} y^A =\epsilon {\dot y}^A,\;\;\;\;
\delta_{\epsilon} b^A =\epsilon\, {\dot b}^A,\;\;\;\;
\delta_{\epsilon} e_i = \frac{d}{d\tau} (\epsilon e_i)\;\;\;\;\; i=1,2
\eqno{(30)}$$
local $b$-rescaling
$$
\delta_{\mu} b^A =\mu \,b^A,\;\;\;\;
\delta_{\mu} y^A =\delta_{\mu} e_i = 0
\eqno{(31)}$$
and reparametrization-like transformation of the form
$$
\delta_{\nu} y^A =\nu p^A, \;\;\;\;
\delta_{\nu} b^A =\frac{\nu}{r^2} \frac{(\dot y, b)}{e_1} y^A,\;\;\;\;
\delta_{\nu} e_1 = \dot \nu,\;\;\;\; \delta_{\nu} e_2 =0
\eqno{(32)}$$
where
$$
p^A = \frac{\dot{y}^A}{e_1} -\frac{e_1^2}{e_2 (\dot{y} , b)^3 }
\bigl\{ \dot{b}^2 +\frac{1}{r^2} (\dot{y},b)^2 \bigr\} b^A.
\eqno{(33)}$$
Here $\epsilon$, $\nu$ and $\mu$ are arbitrary, modulo boundary conditions,
functions of $\tau$.

It should be pointed out that all gauge symmetries (30--32) of the action
(29) remain intact in the case when $y^A$ and $b^A$ are considered to be
$d+1$ vectors in ${\bf R}^{d-1,2} $ subject to Eqs. (2,3).
Therefore, we naturally obtain a model of a spinning particle in
$d$-dimensional AdS space ${\cal M}^d$. The gauge invariance (31) implies that
the configuration space of the model is ${\cal M}^d \times S^{d-2}$.

\section{Physical observables, energy and spin}

The Hamilton generators (17) determine the transformation law of phase-space
functions under the action of the AdS group. On the other hand, they generate
the total set of physical observables of the system. A phase-space function
$\cal F$ is said to be a (strong) physical observable if it commutes with the
first-class constraints
$$
\bigl\{ {\cal F}, T_I \bigr\}_{DB} =0 \;\;\;\;I=5,\ldots ,8
\eqno{(34)}$$
with respect to the Dirac bracket related to the second-class
constraints  $T_1,\ldots ,T_4$:
$$
\{y_A,y_B \}_{DB} = \{y_A,b_B \}_{DB} =\{y_A,k_B \}_{DB} = \{b_A,b_B \}_{DB}
=\{k_A,k_B \}_{DB}=0
$$
$$
\{p_A , p_B \}_{DB} =-\frac{1}{r^2} {\cal J}_{AB}
$$
$$
\{y_A , p_B \}_{DB} =\{b_A , k_B \}_{DB} =\eta_{AB} +\frac{1}{r^2}y_A y_B
\eqno{(35)}
$$
$$
\{ b^A , p^B \}_{DB} =\frac{1}{r^2}y_A b_B\;\;\;\;\;\;
\{ k^A , p^B \}_{DB} =\frac{1}{r^2}y_A k_B.
$$

By direct analysis one can show that any physical observable is a function
of the Hamilton generators only
$$
{\cal F}={\cal F} ({\cal J}_{AB})
\eqno{(36)}$$
on the total surface of constraints.
The same fact follows from more simple consideration. Because of the presence
of four second-class and four first-class constraints, physical phase space
is eight-dimensional. Hence it can be parametrized by $10$ variables
${\cal J}_{AB}$ subject to the constraints $T_7$, $T_8$. As a result, the
Hamilton generators of the AdS group completely specify gauge-invariant
properties of the system.

We turn now to more deep study of the constraints (22,23). It will be shown
that for special values of $\Delta_1$ and $\Delta_2$ the system is
characterized by two physical parameters: {\it minimal positive energy}
$E_o$ and {\it spin} $s$, energy $E$ of the particle being defined as follows
$$
E\equiv {\cal J}^{05} = p^0 y^5 -p^5 y^0 +k^0 b^5 -k^5 b^0 .
\eqno{(37)}$$
Below we will assume  $\Delta_1$ and $\Delta_2$  to be positive, what implies
$p^A$ is a timelike $5$-vector and $k^A$ is a spacelike one.

The fact that the energy (37) might be globally positive definite,
on a connected
component of the constrained surface, lies mainly in the algebraic structure
of the AdS group. For recalling let us consider, for a moment, the model of a
massive spinless particle on the AdS space with the Lagrangian
$$
{\cal L}=\frac{1}{2e}\big({\dot y}^2-\frac{\Delta}{r^2} e^2\big)
\qquad \;\;\;\;\Delta > 0
\eqno{(38)}$$
which is dynamically equivalent to our model in the case $\Delta_2 =0$.
Here the constrained surface is given by
$$
{\cal T}_1 = y^2 +r^2\;\;\;\;\;\; {\cal T}_2 = (y,p)\;\;\;\;\;\;
{\cal T}_3 =r^2 p^2 +\Delta.
\eqno{(39)}$$
and, as a topological space, consists of two connected components determined
by the AdS-covariant conditions
$$
p^0 y^5 -p^5 y^0 >0
\eqno{(40.a)}$$
$$
p^5 y^0 -p^0 y^5 <0.
\eqno{(40.b)}$$
Really, having mutually orthogonal timelike $5$-vectors $p=(p^A)$ and
$y=(y^A)$ one can construct an orthonormal $5$-frame $\{ h_B \}$
$$
h_B =(h^A\,_B) \;\;\;\; (h_B,h_C) = \eta_{BC}
$$
such that $ h_5 =\frac{1}{r}y $, $ h_0 =\frac{r}{\sqrt{\Delta}}p $.
The matrix
$$
h=\bigl( h^A\,_B)
$$
belongs to $O(3,2)$, hence
$$
\bigl| h^5\,_5 h^0\,_0 - h^5\,_0 h^0\,_5 \bigr| \geq 1.
\eqno{(41)}$$
It is now obvious that sign of the energy
$$
{\cal E} = p^0 y^5 -p^5 y^0
\eqno{(42)}$$
is AdS-invariant as well as we have
$$
\bigl| {\cal E} \bigr| \geq \sqrt{\Delta}
\eqno{(43)}$$
at any point of the constrained surface.
Direct computation of $\cal E$ for the phase-space domain
(40.a) leads to the expression
$$
{\cal E}  = \sqrt{\Delta (\frac{\rho}{r})^2 + {\rho}^2 {\vec{p}}\,^2
-(\vec{y},\vec{p})^2}\;.
\eqno{(44)}$$

Similarly to the spinless case, the constrained surface
in our model involves two
connected components specified by the conditions (40). Let us analyse the
function of energy (37) for the points of the component (40.a). It is useful
to express the variables $(p,b,k)$, parametrizing the fiber over some point
$y$ of the AdS space, via those for the fiber over the marked point (6)
$$
(p,b,k) = \cG^{-1} (y) ({\bf p} ,{\bf b} ,{\bf k} )
\eqno{(45)}$$
$\cG (y)$ is given by Eq. (8). We have
$$
{\bf p}^A =(0, {\bf p}^0 ,{\vec{\bf p}} )\;\;\;\;\;\;
{\bf p}^0 =\sqrt{\frac{\Delta_1}{r^2} + {\vec{\bf p}}^2}
$$
$$
{\bf b}^A =(0, {\bf b}^0 ,{\vec{\bf b}} )\;\;\;\;\;\;
\bigl| {\bf b}^0 \bigr| = \bigl| {\vec{\bf b}} \bigr| \neq 0
\eqno{(46)}
$$
$$
{\bf k}^A =(0, {\bf k}^0 ,{\vec{\bf k}} )\;\;\;\;\;\;
{\bf k}^0 =\frac{1}{{\bf b}^0}( {\vec{\bf b}} ,{\vec{\bf k}} )
$$
where we have accounted the constraints. The momenta $p$ and $\bf p$ are
related by the rule
$$
p^5 =-\frac{y^0}{\rho} {\bf p}^0 +\frac{y^5}{r\rho} (\vec{y},
{\vec{\bf p}})
$$
$$
p^0 =\frac{y^5}{\rho} {\bf p}^0 +\frac{y^0}{r\rho} (\vec{y} ,
{\vec{\bf p}})
\eqno{(47)}
$$
$$
\vec{p} ={\vec{\bf p}} +\frac{\vec{y}}{r(\rho+r)} (\vec{y} ,
{\vec{\bf p}})
$$
and similarly for the rest variables. Now, one readily	finds
$$
E = \rho {\bf p}^0 -\frac{1}{r} {\bf b}^0 (\vec{y}, \vec{w} )
\eqno{(48)}$$
where
$$
\vec{w} =  {\vec{\bf k}} -{\vec{\bf b}} \frac {({\vec{\bf b}},
{\vec{\bf k}} )}{{\vec{\bf b}}^2}.
\eqno{(49.a)}$$
It is important that the $3$-vector just introduced possesses the properties
$$
(\vec{w},{\vec{\bf b}})=0\;\;\;\;\;\;\vec{w}^2 ={\bf k}{}^2 =k^2.
\eqno{(49.b)}$$
Then, further use of the constraints allows to get the following unequality
$$
\left| \frac{1}{r} {\bf b}^0 (\vec{y},\vec{w}) \right|
\leq \frac{\sqrt{\Delta_2}}{\Delta_1} |\vec{y}| ({\bf p}^0 +
|{\vec{\bf p}|})
\eqno{(50)}$$
and hence
$$
E\geq \rho {\bf p}^0 -
\frac{\sqrt{\Delta_2}}{\Delta_1} |\vec{y}| ({\bf p}^0 +
|{\vec{\bf p}}|)  .
\eqno{(51)}$$
The expressions in both sides coincide under the conditions
$$
\vec{y}\parallel\vec{w}\;\;\;\;\;\;\;\;{\vec{\bf p}}\parallel {\vec{\bf b}}
\;\;\;\;\;\;\;\;{\bf b}^0 >0
\eqno{(52.a)}$$
and therefore
$$
\vec{y}\perp\vec{b}\;\;\;\;\;\;\;\;\;\;\vec{p}\parallel \vec{b}
\eqno{(52.b)}$$
as a consequence of Eqs. (47), (49.b). Here we have used the fact
that the energy is invariant under the transformations ${\bf b}
\rightarrow \lambda{\bf b}$, ${\bf k}\rightarrow \lambda^{-1}{\bf k}$,
hence ${\bf b}^0$ can be chosen to be positive.

Because of Eq. (51), the lower boundary of values of the energy
$E=E(y,p,b,k)$ is given by the function
$$
f(u,v) = \sqrt{(r^2 + u^2) (\frac{\Delta_1}{r^2} +v^2) }
-\frac{\sqrt{\Delta_2}}{\Delta_1}
u \left( v + \sqrt{\frac{\Delta_1}{r^2} +v^2}\;\right)
\eqno{(53.a)}
$$
$$
u=|\vec{y}|\;\;\;v=|\vec{p}|\;\;\;\;\;\;\;\;\;\;\;\;\; 0\leq u,v \leq\infty.
\eqno{(53.b)}
$$
This function proves to be positive definite if and only if
$$
2\sqrt{\Delta_2} \leq \Delta_1 .
\eqno{(54)}
$$
Thus the case  $\Delta_1 <2\sqrt{\Delta_2}$ is unphysical, for
the energy can take zero and negative values, and should be ruled out
from our consideration.

The choice
$$
2\sqrt{\Delta_2} = \Delta_1
\eqno{(55)}$$
is very special, since the
energy can sink down to zero in a limit of infinite $|\vec{y}|$ and
$|\vec{p}|$. A similar situation takes place for the massless spinless
particle (see Eq. (44) with $\Delta =0$); here the energy is positive
definite but can approach zero for $|\vec{p}|
\rightarrow 0$. Finally, the case
$$
2\sqrt{\Delta_2} < \Delta_1
\eqno{(56)}$$
is characterized by the property
$$
\lim\limits_{ {\displaystyle u,v\to +\infty } } f(u,v) = +\infty .
\eqno{(57)}$$
The same behaviour is characteristic of the massive spinless particle,
in accordance with Eq. (44). Thus we are tempted to treat the choices
(55) and (56) to be characteristic of massless and massive spinning particles,
respectively.
${}$From now on, we restrict ourselves to the consideration of the case (56).

Introducing new parameters $E_o$ and $s$ defined by
$$
E_o^2 \equiv \frac{\Delta_1}{2} + \sqrt{\left(\frac{\Delta_1}{2}\right)
^2-\Delta_2}\;\;\;\;\;\; \;\;\;
s^2 \equiv \frac{\Delta_1}{2} - \sqrt{\left(\frac{\Delta_1}{2}\right)^2-
\Delta_2}
\eqno{(58)}$$
Eq. (56) is seen to be equivalent to the relation
$$
s<E_o
\eqno{(59)}$$
and the original parameters are expressed as follows
$$
\Delta_1=E_o^2+s^2
\eqno{(60.a)}$$
$$
\Delta_2=E_o^2 s^2.
\eqno{(60.b)}$$
In the domain of non-zero $u$ and $v$, $f$ proves to possess the only
extremal point
$$
\mid \vec{y} \mid=\frac{rs}{\sqrt{E_o^2-s^2}}
\eqno{(61.a)} $$

$$
\mid \vec{p} \mid=\frac{s^2}{r\sqrt{E_o^2-s^2}}
\eqno{(61.b)}$$
in which $f$ is equal to $E_o$. Then, Eq. (57) and the obvious relations
$$
E_o<f(0,v)\;\;\;\;\;\;E_o<f(u,0)
\eqno{(62)}$$
imply that $E_o$ is the global minimum of the energy
$$
E \geq E_o.
\eqno{(63)}$$
The global minimum is achieved in those points
of the constrained surface
which obey the relations (52) and (61). In every such point, the Hamilton
generators turn out to have the form
$$
{\cal J}^{AB}=\left(\begin{array}{cccc}
\displaystyle 0 & \displaystyle{-E_o} & \vdots & \displaystyle 0\\
\displaystyle{E_o} & \displaystyle 0 & \vdots & \displaystyle 0\\
\dotfill & \dotfill & \vdots & \dotfill\\
\displaystyle 0 &\displaystyle 0 &\vdots & {\cal J}^{ij}\end{array}\right)
\;\;\;\;\;\quad \sum_{i<j}^{} \bigl ({\cal J}^{ij} \bigr)^2=s^2
\eqno{(64.a)}$$
and the classical Pauli-Lubanski vector (15) reads
$$
W^A=(0,0,W^i)\;\;\;\;\;\quad W^i=\frac{1}{2}E_o\varepsilon^{ijk}{\cal J}^{jk}.
\eqno{(64.b)}$$
Thus, the parameter $s$ has the sence of the total angular momentum of a
particle with the minimal energy. That is why we can identify $s$
with spin.

It should be remarked that the massive spinning particle having the
minimal energy remains to stay at the finite distance (61.a) from the origion
$\vec{y}=0$ and its ``3-momentum'' has the constant non-zero value (61.b).
The particle moves along a circle, with center at the origion,
that lies in the plane orthogonal to $\vec{W}$ (64.b). This picture
drastically differs from that for the massive spinless particle
which gets the minimum of the energy only when $\vec{y}=0$  and $\vec{p}=0$.
In accordance with Eq. (61), the dynamical behaviour of the spinning
particle looks similar to that of the spinless one only when $E_o \gg s$.
For $E_o \approx s$, however, the spin effects become very strong
and cause the particle with $E=E_o$ to be located far away from the origion.
Another important remark is that the conditions (61), characterizing the
states with the minimal energy, are invariant with respect to the
gauge transformations induced by the constraints.

\section{Spinning particle and inner AdS-geometry}

In this section, the model will be reformulated as a dynamical system
on a curved space. We start with resolving the second-class constraints
(18,19) and the Dirac bracket (35) via unconstrained variables on the
cotangent bundle of $T(M^4)$ and related geometric objects.
Then we describe the model in terms of inner coordinates on ${\cal M}^6$.

Let $x^m$, $m$=0,1,2,3, be local coordinates on the surface (1). The
induced metric ${\rm d}s^2=\eta_{AB}{\rm d}y^{A}{\rm d}y^{B}$ reads
$$
{\rm d}s^2=g_{mn}(x){\rm d}x^{m}{\rm d}x^{n}
\eqno{(65)}$$
$g_{mn}$ being a metric of constrant negative curvature ${\cal R}=-12/r^2$.
Now the cotangent bundle of $T(M^4)$ can be parametrized by unconstrained
variables $(x^m , {\bf p}^a ,{\bf b}^a ,{\bf k}^a )$, where
${\bf p}$, ${\bf b}$ and ${\bf k}$ are defined by the rule (45),
for ${\cal G}(y)$  a solution of Eqs. (5,6). Obviously,
${\bf p}$, ${\bf b}$ and ${\bf k}$ transform as 4-vectors with respect
to the local Lorentz group and as scalars under the general covariance group
of $M^4$. Associated with ${\cal G}(y)$ is a vierbein $e_m\,^a (x)$
of the metric that converts curved-space indices into flat-space ones.
Really, let us define
$$
e_m\,^A \equiv {\cal G}^A\,_B \frac{\partial y^B}{\partial x^m} =
-\frac{\partial {\cal G}^A\,_B}{\partial x^m}y^B = (0, e_m\,^a)
\eqno{(66)}$$
where we have used the identity
$$
{\cal G}^5\,_B =-\frac{1}{r}y_B.
\eqno{(67)}$$
Since ${\cal G}(y)$ belongs to $O(3,2)$, one readily gets the relations
$$
g_{mn}={e_m}^{a}{e_n}^{b}\eta_{ab}.
\eqno{(68)}$$

By construction, the functions $x(y)$ and ${\cal G}(y)$ are defined only
on the AdS hyperboloid. They can be uniquely extended onto the subspace of
${\bf R}^{3,2}$ [13]
$$
{\cal W}=\{y \in {\bf R}^{3,2},\;\;\; y^2<0 \}
\eqno{(69)}$$
if one restricts them to have zeroth order of homogeneity in $y$
$$
\frac{\partial x^m}{\partial y^C}y^C =0   \;\;\;\;\;\;\;
\frac{\partial {\cal G}^A\,_B}{\partial y^C}y^C =0.
\eqno{(70)}$$
Thus the variables $x^m$ and $\sigma$, $\sigma\equiv (-y^A y_A)^{-1/2}$,
can be chosen to parametrize $\cal W$
instead of $y^A$ (note, $\sigma$ commutes
with any phase-space function  with respect to the Dirac bracket (35)).
Introducing
$$
e_A\,^m \equiv \frac{\partial x^m} {\partial y^B} ({\cal G}^{-1})^B\,_A  =
(0, e_a\,^m)
\eqno{(71)}$$
one finds
$$
e_a\,^m  e_m\,^b =\delta_a^b
\eqno{(72)}$$
therefore $e_a\,^m (x)$ is the inverse vierbein.

Another geometric object,
the torsion-free spin connection
$\omega_{mab}(x)$, defined by
$$
\omega_{mab}= -\omega_{mba} \;\;\;\;\;\;\;\;
{T_{mn}}^a = \partial_n {e_m}^a - \partial_m {e_n}^a + {\omega_n}^a\,_b
{e_m}^b - {\omega_m}^a\,_b {e_n}^b =0
$$
can be represented
in the form
$$
{{\omega_m}^a}_b={{\cal G}^a}_C\frac{\pa ({{\cal G}^{-1})^C}_{b}}{\pa x^m}.
\eqno{(73)}$$

Let us pass from ${\bf p}_a$ to new variables $p_m$ defined as follows
$$
{\bf p}_a =e_a\,^m {\bf p}_m \;\;\;\; {\bf p}_m=p_m-\frac{1}{2}
\omega_{mcd} {\cal M}^{cd}
\eqno{(74)}$$
where
$$
{\cal M}^{cd} = {\bf k}^c {\bf b}^d -{\bf k}^d {\bf b}^c.
\eqno{(75)}$$
It is now an instructive exercise to verify, with the help of
Eqs. (66--73), that
the Dirac brackets (35) are equivalent to the canonical commutation relations
for the variables $x^m$, $p_n$, ${\bf b}^a$ and ${\bf k}_b$ :
$$
\{x^m , p_n \}_{DB}=\delta^m_n\;\;\;\;\;\;\;
\{ {\bf b}^a ,{\bf k}_b \}_{DB} =\delta^a_b
\eqno{(76)}$$
the other brackets vanish. Eq. (75) defines the Hamilton generators
of the Lorentz transformations acting on ${\bf b}^a$ and ${\bf k}_b$
and leaving $x^m$ and $p_n$ inert. In the phase-space variables inroduced,
the second class constraints are completely eliminated and we rest with
the first-class constraints (20--21). Finally, the Lagrangian (29)
takes the form
$$
{\cal L}=\frac{1}{2e_1}\bigl( g_{mn}\dx^m\dx^n-\frac{\Delta_1}{r^2}
e_1^2 \bigr)+
\frac{1}{2e_2}\Biggl(\frac{\stackrel{\bullet}{{\bf b}}\!^a
\stackrel{\bullet}{{\bf b}}\!_a}
{(\dx{}^me_m\,^c {\bf b}_c)^2}e_1^2 +
\frac{\Delta_2}{r^2} e_2^2 \Biggr)
\eqno{(77)}$$
where
$$
\stackrel{\bullet}{{\bf b}}\!^a=\dot{{\bf b}}^a +\dot{x}^m {\omega_m}
^a\,_c {\bf b}^c
\eqno{(78)}$$
is the Lorentz-covariant derivative of ${\bf b}^a$ along the world-line.
Here ${\bf b}^a$ is constrained to be a lightlike 4-vector.

The above Lagrangian is a minimal curved-space extension of that recently
proposed in Ref. [6] for the massive spinning particle in Minkowski space:
$$
{\cal L}_{flat}=\frac{1}{2e_1}\bigl( \dx^a\dx_a-(me_1^2) \bigr)+
\frac{1}{2e_2}\bigg(\frac{\dot{{\bf b}}^a \dot{{\bf b}}_a}
{(\dx{}^c {\bf b}_c)^2}e_1^2 +
(mse_2)^2 \bigg).
\eqno{(79)}$$

The action functional in Minkowski space possesses three types of local
symmetries which can be read off from Eqs. (30--33) by making obvious
replacements and setting $r^{-2} =0$. The Lagrangian (77)
is formally well-defined
not only for the AdS space but for arbitrary curved space-time. However,
choosing in (77) $e_m\,^a (x)$ to be the vierbein of a space-time with
non-constant curvature ${\cal R}_{mnab}$, the action turns out to be invariant
only under the general coordinate transformations and local $b$-rescalings.
The existence of three local symmetries takes place if and only if the
curvature of space-time is constant. This can be seen as follows.
The ``covariant derivatives'' (74) satisfy the commutation
relations
$$
\{{\bf p}_m,{\bf p}_n\}=-\frac{1}{r^2}e_m\,^a e_n\,^b {\cal M}_{ab}
\eqno{(80)}$$
in the AdS space and
$$
\{{\bf p}_m,{\bf p}_n\}=\frac{1}{2} {\cal R}_{mn}\,^{ab} {\cal M}_{ab}
\eqno{(81)}$$
in general case. For the constraints $T_7$ and $T_8$  (22,23,26) rewritten
in the form
$$
\frac{1}{r^2}T_7= g^{mn} {\bf p}_m {\bf p}_n
+\frac{\Delta_1}{r^2} \approx 0
\eqno{(82.a)}
$$
$$
\frac{1}{r^2}T_8=({\bf b}^a e_a\,^m {\bf p}_m)^2 {\bf k}^c
{\bf k}_c - \frac{\Delta_2}{r^2} \approx 0
\eqno{(82.b)}
$$
Eqs. (76), (81) give
$$
\frac{1}{r^4} \{ T_7 , T_8 \} = 2({\bf p},{\bf b})
{\bf k}^2 {\bf b}^a {\bf p}^b {\cal R}_{abcd} {\cal M}^{cd}.
\eqno{(83)}$$
The expression in the r.h.s of (83) vanishes on the surface of constraints
$T_5$, $T_6$ if and only if
$$
{\cal R}_{abcd} = \frac{1}{12}{\cal R}(\eta_{ac}\eta_{bd}-\eta_{ad}\eta_{bc})
$$
where the scalar curvature $\cal R$ is a constant, as a consequence of the
Bianchi identities. Therefore, the constraints $T_5,\ldots,T_8$ can form
a first-class algebra only in the case of space-times with constant curvature.

Now, let us reformulate the model in terms of internal variables on
${\cal M}^6$.
We standardly cover $S^2 = {\bf C} \cup \{ \infty \}$, considered
as a one-dimensional complex manifold, by two open
patches $U_1 = {\bf C}$ and $U_2 = {\bf C}^{*} \cup \{ \infty \}$
parametrized by complex coordinates $z$ and $\omega$, respectively,
which are related in the overlap $U_1 \cap U_2 = {\bf C}^*$ by the
rule $\omega = -1/z$.

The Lorentz group $SO^{\uparrow}(3,1)\cong SL(2,{\bf C}) / {\bf Z}_2$
acts on $S^2$ by
fractional linear transformations of the form
$$
z' = \frac{az-b}{-cz+d} \qquad \|N\|={N_\alpha}^\beta=\left(
\begin{array}{cc} a & b\\ c & d\end{array}\right)\in SL(2,{\bf C}).
\eqno{(84)}$$
This implies that the two-component objects
$$
z^\alpha= (z)^\alpha =(1,z) \qquad \;\;\;\alpha =0,1
$$
$$
\omega_\alpha= (\omega)^\alpha =(1,\omega) =
- z^{-1}\;\varepsilon_{\alpha\beta} z^{\beta}
\eqno{(85)}$$
transform under the action of the Lorentz group (84)
simultaneously as left Weyl spinors and spinor fields on $S^2$:
$$
{z'}^\alpha =\left(\frac{\pa z'}{\pa z}\right)^{1/2} z^\beta
(N^{-1})_\beta{}^\alpha
$$
$$
{\omega'}_\alpha =\left(\frac{\pa \omega'}{\pa \omega}\right)^{1/2}
(N)_\alpha{}^\beta \omega_\beta.
\eqno{(86)}$$
Our two-component spinor notations and conventions coincide with those
adopted in [22], in particular, $\varepsilon_{\alpha\beta}$ is the
spinor metric.

The lightlike variable ${\bf b}^a$, entering the Lagrangian (77), can be
parametrized as follows
$$
{\bf b}^a = \xi^a \varphi({\bf b}) = \zeta^a \psi({\bf b})
\eqno{(87)}$$
where
$$
\xi^a = (\sigma^a)_{\alpha\dot\alpha}z^\alpha{\bar z}^{\dot\alpha}
$$
$$
\zeta^a = (\sigma^a)_{\alpha\dot\alpha}\omega^\alpha
{\bar \omega}^{\dot\alpha}
\eqno{(88)}$$
and
$$
\psi({\bf b}) = |\omega|^2 \varphi({\bf b})
$$
for some function $\varphi({\bf b})$.
Here $(\sigma^a)_{\alpha\dot\alpha} $
are the Pauli matrices [22], ${\bar z}^{\dot\alpha}$ and
${\bar \omega}^{\dot\alpha}$ are the complex conjugates of $z^\alpha$ and
$\omega^\alpha$, respectively. Then, with the use of the identity
$$
\varepsilon^{\alpha\beta}=z^\alpha\pa_zz^\beta -z^\beta\pa_zz^\alpha
\eqno{(89)}$$
the Lagrangian (77) can be rewritten in the form
$$
{\cal L}=\frac{1}{2e_1}\bigl( g_{mn}\dx^m\dx^n-\frac{\Delta_1}{r^2}
e_1^2 \bigr)+
\frac{1}{2e_2}\Biggl(\frac{4\stackrel{\bullet}{|z|}\!^2}
{(\dx{}^me_m\,^a \xi_a)^2}e_1^2 +
\frac{\Delta_2}{r^2} e_2^2 \Biggr)
\eqno{(90)}$$
where
$$
\stackrel{\bullet}{z} =
\dot{z} +\frac{1}{2}\dot{x}^m\omega_{mab}(x)\Sigma^{ab}
\eqno{(91)}$$
$$
\Sigma^{ab} = - (\sigma^{ab})_{\alpha\beta}z^\alpha z^\beta
\eqno{(92)}$$
with $\sigma^{ab}$ the spin matrices [22]. Contrary to $\dot z$,
$\stackrel{\bullet}{z}$ is changed homogeneously with respect to
the local Lorentz transformations.

The multipliers $e_1$ and $e_2$ can be eliminated with the aid of their
equations of motion. Then $\cal L$ turns into
$$
{\cal L}=-\frac{1}{r} \sqrt{
-g_{mn}\dx^m\dx^n \Biggl(\Delta_1-2\sqrt{\Delta_2}
\frac{2r\stackrel{\bullet}{|z|}}
{|\dx{}^me_m\,^a \xi_a|}  \Biggr)}\;.
\eqno{(93)}$$
The Lagrangian is well defined if the admissible values of particle's
velocity in the internal space $S^2$ are constrained on the manner
$$
\Delta_1 > 2\sqrt{\Delta_2} \frac{2r\stackrel{\bullet}{|z|}}
	   {|\dx{}^me_m\,^a \xi_a|}
$$
in addition to the causal restriction $g_{mn}\dx^m\dx^n < 0$ in the
space-time.

%------------------------------------------------------------------------
\section{Dynamical histories}

We finally turn to the analysis of dynamical histories in the model.
To completely describe the mass shell, it is sufficient in fact to
determine solutions of the dynamical equations with the minimal energy.
The trajectories with $E > E_o$ can be simply restored from those with
$E=E_o$ by applying all possible AdS-transformations.

The Hamiltonian action leading to the Lagrangian (77) reads
$$
S=\int{\rm d}\tau \left(p_m \dot x^m+{\bf k}_a \dot{{\bf b}}{}^a-\frac{1}{2}
\sum_{I=5}^8 \lambda_I T_I\right).
\eqno{(94)}$$
where $T_5={\bf b}^2$, $T_6=({\bf b},{\bf k})$, and the constraints $T_7,T_8$
are given by Eqs. (82,74,75). The corresponding equations of motion
can be represented, with the aid of the constraints, in the form
$$
\dot{x}^m - r^2\lambda_7{\bf p}^m - \Delta_2 \lambda_8 \frac{{\bf b}^m}
{({\bf b},{\bf p})} = 0
\eqno{(95.a)}$$
$$
\stackrel{\bullet}{{\bf p}}\!_a - \frac{1}{2}{\cal R}_{abcd}\dot{x}^m
e_m\,^bM^{cd} =0
\eqno{(95.b)}$$
$$
\stackrel{\bullet}{{\bf b}}\!^a - \frac{1}{2}\lambda_6 {\bf b}{}^a
- r^2\lambda_8 ({\bf b},{\bf p})^2 {\bf k}^a = 0
\eqno{(95.c)}$$
$$
\stackrel{\bullet}{{\bf k}}\!^a + \lambda_5 {\bf b}^a + \frac{1}{2}\lambda_6
{\bf k}^a + \Delta_2 \lambda_8 \frac{{\bf p}^a}{({\bf b},{\bf p})} = 0 .
\eqno{(95.d)}$$
Here the covariant differentiation is defined by the rule (78),
${\cal R}_{abcd}$ is the curvature tensor of the AdS space
$$
{\cal R}_{abcd} = - \frac{1}{r^2}(\eta_{ac} \eta_{bd} - \eta_{ad}\eta_{bc}).
$$
The equations of motion should be supplemented by the global restriction
on Lagrange multipliers
$$
\Delta_1 \lambda_7 > \Delta_2 \lambda_8
\eqno{(96)}$$
that guarantees the timelikeness of $\dot{x}^m$ at any point of the world
line.

The above action possesses four gauge symmetries related to the constraints.
Those associated with $T_5$ and $T_6$ can be fixed by imposing the
following AdS-invariant gauge conditions
$$
({\bf p},{\bf k}) = 0
\eqno{(97.a)}$$
$$
({\bf p},{\bf b}) = -1 .
\eqno{(97.b)}$$
Then on the mass shell, $\lambda_5$ and $\lambda_6$ are completely
determined in the form
$$
\lambda_5 = -\frac{\Delta_2}{r^2}(\lambda_7 -\Delta_1 \lambda_8)
\qquad \lambda_6 = 0.
\eqno{(98)}$$
Importantly enough, Eq. (97.a) implies
$$
{\bf k}^0 = 0
\eqno{(99)}$$
for any dynamical history with the minimal energy.

Let us now choose a useful coordinate system $\{x^m\}$ on $M^4$ as follows
$$
y^5 = \rho \cos \bigl(x^0 / r \bigr)
$$
$$
y^0 = \rho \sin \bigl(x^0 / r \bigr)
\eqno{(100)}$$
$$
\vec{y} =\vec{x} .
$$
Here $0\leq x^0 < 2\pi r$ for $M^4$ and $-\infty < x^0 < \infty$ for
the universal covering space of $M^4$. We also set ${\cal G}(y)$ up as in
Eq. (8). Then the vierbein reads
$$
e_m\,^a=\left(\begin{array}{ccc} \frac{{\displaystyle{\rho}}}
{{\displaystyle{r}}} & \vdots & 0\\
\dotfill & \vdots & \dotfill\\
\displaystyle 0 & \vdots &
\delta^{ij} - \frac{{\displaystyle{x^i x^j}}}
{{\displaystyle{\rho (\rho + r)}}} \end{array}\right) \qquad
i,j=1,2,3
\eqno{(101)}$$
and the spin connection $w_{abc} = e_a\,^m w_{mbc}$ looks like
$$
w_{0i0} = -w_{00i} =\frac{x_i}{r\rho} \qquad w_{i0j} = 0
$$
$$
w_{ijk} = \frac{1}{r(\rho + r)}(\delta_{ij}x_k - \delta_{ik}x_j )
\eqno{(102)}$$
where $i,j,k =1,2,3$.

For the trajectories with the minimal energy, the equations of motion are
drastically simplified. The relations (52,98) and gauge conditions (97)
imply that in this case
the only time-dependent functions to be determined are $x_0(\tau)$,
$\vec{m}(\tau)$ and $\vec{n}(\tau)$, where
$$
\vec{m} = \frac{\vec{y}}{\bigl |\vec{y} \bigr |} =
\frac{\vec{{\bf k}}}{\bigl |\vec{{\bf k}} \bigr |} \qquad
\vec{n} = \frac{\vec{{\bf p}}}{\bigl |\vec{{\bf p}} \bigr |} =
\frac{\vec{{\bf b}}}{\bigl |\vec{{\bf b}} \bigr |}
\eqno{(103)}$$
It is natural here to recall that the 3-vectors $\vec m$, $\vec n$ and
$\vec W$ (64.b) form an orthonormal set at each moment of time evolution,
$\vec W$ being time-independent. With the use of the expressions (101,102),
we derive from (95) the following equations
$$
\frac{\dot{x}^0}{r} = E_o(\lambda_7 - s^2\lambda_8)
\eqno{(104)}$$
$$
\dot{\vec{m}} = s\vec{n}(\lambda_7 - E_o^2\lambda_8)
\eqno{(105.a)}$$
$$
\dot{\vec{n}} = -  s\vec{m}(\lambda_7 - E_o^2\lambda_8).
\eqno{(105.b)}$$

It follows from Eq. (96) that the r.h.s in (104) is strictly positive
at each point of the world line
$$
\lambda_7 - s^2\lambda_8 > 0.
\eqno{(106)}$$
Thus we are able to fix the gauge freedom associated with $T_7$ by imposing
the gauge condition
$$
x^0 = r\tau
\eqno{(107)}$$
which relates the evolution parameter to the physical time.
There is no physically preferable way to fix the gauge freedom related to
$T_8$. The most elegant gauge condition seems to be
$$
\lambda_8 = 0.
\eqno{(108)}$$
Then on the mass shell, ${\bf b}$ is covariantly constant modulo local
$\bf b$-rescalings, in accordance with (95.c), and strictly covariantly
constant in the gauge (97). Choosing the gauge conditions (107,108),
the equations (105) take the form
$$
\dot{\vec{m}} = \frac{s}{E_o}\vec{n} \qquad
\dot{\vec{n}} = - \frac{s}{E_o}\vec{m} .
\eqno{(109)}$$
In this gauge, the trajectories for $s \neq 0$ are in general globally
defined only on the universal
covering space of $M^4$. The same feature is known for the on-shell fields
describing free particles with spin on the AdS space [9,13].

%%%%%%%%%%%%%%%%%%%%%%%%%%%%%%%%%%%%%%%%%%%%%%%%%%%%%%%%%%%%%%%%%%%%%%%%%%%%
%%%%%%%%%%%%%%%%%%%%%%%%%%%%%%%%%%%%%%%%%%%%%%%%%%%%%%%%%%%%%%%%%%%%%%%%%%%%
%%%%%%%%%%%%%%%%%%%%%%%%%%%%%%%%%%%%%%%%%%%%%%%%%%%%%%%%%%%%%%%%%%%%%%%%%%%%%
\section{Conclusion }

Let us give a brief overview of the results and some comments.
In this paper we have
suggested the model for a spinning particle on $d$-dimensional
anti-de Sitter space as a mechanical system with
the configuration space
${\cal M}^{2(d-1)} = M^d \times S^{d-2}$.
In any space-time dimensions,
the model possesses two gauge symmetries. Their origion lies in the
fact that the phase-space counterparts of the second- and fourth-order
Casimir operators of $so(d-1,2)$ are constrained to have constant values
$\Delta_1$, $\Delta_2$ on the phase space.

We have thoroughly studied the model in four dimensions.
For $\Delta_2 =0 < \Delta_1$, our model is equivalent to the theory
of a massive spinless particle (38). The case of massive spinning particles
is specified by the condition
$0 < 2\sqrt{\Delta_2} < \Delta_1$. Only under the choice
$0 \leq 2\sqrt{\Delta_2} < \Delta_1$, the energy proves to have
a global positive minimum $E_o$ over the phase space, $E_o$ given by Eq.
(58). The last condition appears to be equivalent to the relation
$E_o > s \geq 0$, where $s$ (58) is the value of the total angular
momentum in any state with the minimal energy. Thus, $s$ can be naturally
identified with the spin of the particle. It is
worth noting that the inequality
$E_o > s$ has been postulated in Ref. [20] to be the classical analogue
of the quantum massive condition $E_o > s + 1$.

In summary, in $d=4$ and for the restricted set of parameters
$0 \leq 2\sqrt{\Delta_2} < \Delta_1$, our model can be conceptually
treated as a minimal and unified model of a massive particle
with fixed, but arbitrary, spin on the AdS space.
The model is minimal, because its configuration manifold has minimally
possible dimension to describe the evolution in space-time
and spin dynamics. It is unified, since the configuration
space is spin-independent; the dynamics is specified by the choice
of parameters $\Delta_1$, $\Delta_2$ entering the Lagrangian.

When $2\sqrt{\Delta_2} = \Delta_1$, the energy remains positive over
the phase space, but has no global minimum and can approach zero in some
limiting directions. This choice of the parameters should correspond
to massless particles. However, a simple counting shows that for
$\Delta_2 > 0$ the model has too many degrees of freedom to describe
a true massless dynamics. It is still unclear whether there exists
a universal configuration space for massless spinning particles or not.

As we have demonstrated in sec. 6, the dynamical equations of the model
can be readily integrated after specifying simple gauge conditions.
But since the original equations of motion (95) involve arbitrary
functions $\lambda_5,\ldots,\lambda_8$, which get fixed only upon
putting gauge conditions up, the explicit structure of dynamical
trajectories is gauge dependent and hence has no direct physical interest.
Among gauge invariant properties of the model are the conditions (61) that
characterize the states with the minimal energy. Eq. (61.a) leads to
a rather unusual effect, from the point of view of our flat-space
intuition. It is seen that when $E_o$ approaches its lower bounbary,
$E_o \approx s$, the particle with $E = E_o$ turns out to be located
far away from the origion $\vec{y} = 0$. Nevertheless, this result
is very natural for the AdS universe and can be explained similar to what
have been said (see, for example, [11]) to demonstrate the statement
that singleton are physically unobservable. Indeed, let us pass from
dimensionless units for energy to standard ones by redefining $E \rightarrow
r^{-1}E$. Then the relation $E_o \approx s$ turns into $energy =
r^{-1} \times angular$ $momentum$, $r^{-2}$ being proportional to the
curvature of the AdS space and hence very small. Therefore, the particle
moves at distances of the same order as the ``radius''of the AdS space.

Finally, let us shortly comment on quantization of the model. It has
been argued, in sec. 4, that every physical observable in the model
is a function of the Hamilton generators of $so(3,2)$. So, the covariant
operatorial quantization of the model reduces to constructing unitary,
positive-energy representatios of $so(3,2)$, what has been
worked out in Ref. [10]. A non-trivial question, however, is how  to
construct a coordinate realization for this (constrained) quantum
mechanics, i.e. to realize  the massive, positive-energy
AdS-representations in some function spaces over ${\cal M}^6$
with an apropriate Hilbert space structure. In the case of flat space-time,
the analogous problem has been exhaustively studied in our previous paper
[6] where all such realizations were classified. We intend to give a similar
consideration for the AdS case in a forthcoming publication [21].

\vspace{0.5cm}

\noindent
{\bf Acknowledgements} \\
\noindent
It is a pleasure to thank N. Dragon, O. Lechtenfeld and A. Zubarev
for useful discussions.
This work was supported in part by European
Community grant No. INTAS-93-2058 and by the International Science
Foundation grant No. M21000. The work of SMK was supported in part
by the Alexander von Humboldt Foundation.
The work of SLL was partly supported by the Royal Society under
a Kapitza Fellowship programme.

\end{document}